\documentclass[a4paper,pre,twocolumn,amsmath,amssymb,longbibliography]{revtex4-1}
\usepackage{bm}
\usepackage{graphicx,color,soul}
\usepackage[colorlinks,linkcolor=magenta,citecolor=magenta]{hyperref}

\newcommand{\dd}{\mathrm{d}}

\newcommand{\fd}[2]{\frac{\delta #1}{\delta #2}}
\newcommand{\mean}[1]{\langle #1 \rangle}
\newcommand{\Int}[1]{\int\dd #1\;}
\newcommand{\IInt}[3]{\int_{#2}^{#3}\dd #1\;}
\newcommand{\IntF}[1]{\int\frac{\dd^d #1}{(2\pi)^d}\;}

\renewcommand{\vec}[1]{\mathbf #1}

\newcommand{\al}{\alpha}
\newcommand{\gam}{\gamma}

\newcommand{\eps}{\varepsilon}
\newcommand{\kap}{\kappa}
\newcommand{\lam}{\lambda}
\newcommand{\Lam}{\Lambda}

\newcommand{\sig}{\sigma}

\newcommand{\om}{\omega}
\newcommand{\Om}{\Omega}

\newcommand{\tx}{\tau_\text{r}}
\newcommand{\x}{\vec r}

\newcommand{\im}{\text{i}}
\newcommand{\rhoc}{\rho_\text{c}}
\newcommand{\ellc}{\ell_\text{c}}

\begin{document}

\title{Critical behavior of active Brownian particles: Connection to field theories}

\author{Thomas Speck}
\affiliation{Institut f\"ur Physik, Johannes Gutenberg-Universit\"at Mainz, Staudingerweg 7-9, 55128 Mainz, Germany}


\begin{abstract}
  We explore the relation between active Brownian particles, a minimal particle-based model for active matter, and scalar field theories. Both show a liquid-gas-like phase transition towards stable coexistence of a dense liquid with a dilute active gas that terminates in a critical point. However, a comprehensive mapping between the particle-based model parameters and the effective coefficients governing the field theories has not been established yet. We discuss conflicting recent numerical results for the critical exponents of active Brownian particles in two dimensions. Starting from the intermediate effective hydrodynamic equations, we then present a novel construction for a scalar order parameter for active Brownian particles that yields the ``active model B+''. We argue that a crucial ingredient is the coupling between density and polarization in the particle current. The renormalization flow close to two dimensions exhibits a pair of perturbative fixed points that limit the attractive basin of the Wilson-Fisher fixed point, with the perspective that the critical behavior of active Brownian particles in two dimensions is governed by a strong-coupling fixed point different from Wilson-Fisher and not necessarily corresponding to Ising universality.
\end{abstract}

\maketitle


\section{Introduction}

The comprehensive understanding of critical phenomena has been one of the major successes of statistical mechanics~\cite{ma18}, which has cumulated in the renormalization group theory and the Nobel prize for Kenneth Wilson~\cite{wilson83}. While pushed forward by continuous transitions in equilibrium systems, critical behavior is not restricted to thermal equilibrium and many driven systems exhibit special points and phases that are governed by scale invariance~\cite{hinrichsen00,odor04,binder21}. One class of steadily driven systems in particular, active matter comprised of many interacting ``units'' undergoing directed motion (ranging from birds to bacteria to synthetic colloidal particles), has received extensive attention lately~\cite{bechinger16,fodor18,gompper20}. Among others, the emergence of spontaneous polar order~\cite{toner95,toner98}, active nematics~\cite{mishra10}, and the self-organization of chemotactic cells and bacteria~\cite{gelimson15,mahdisoltani21} have been studied through the lens of critical phenomena and the dynamic renormalization group~\cite{forster77,medina89}.

Within the wider field of active matter, active Brownian particles (ABPs) have become an intensively studied paradigm for the interplay of directed motion with short-range repulsive forces modeling volume exclusion~\cite{bialke15a}. ABPs are propelled through adding a velocity term $v_0\vec e$ with fixed magnitude $v_0$ and freely diffusing, \emph{independent} unit orientations $\vec e$ (no alignment). As first shown in simulations~\cite{fily12}, this model undergoes \emph{motility-induced phase separation}~\cite{cates15} that resembles liquid-gas phase separation in a passive fluid (or suspension) but in the absence of cohesive forces. Instead, the blocking of particles due to the repulsive forces in combination with the persistence of forward motion leads to a dynamic instability through the dependence of the effective propulsion speed on the local density~\cite{bialke13}.

\begin{figure}[b!]
  \centering
  \includegraphics{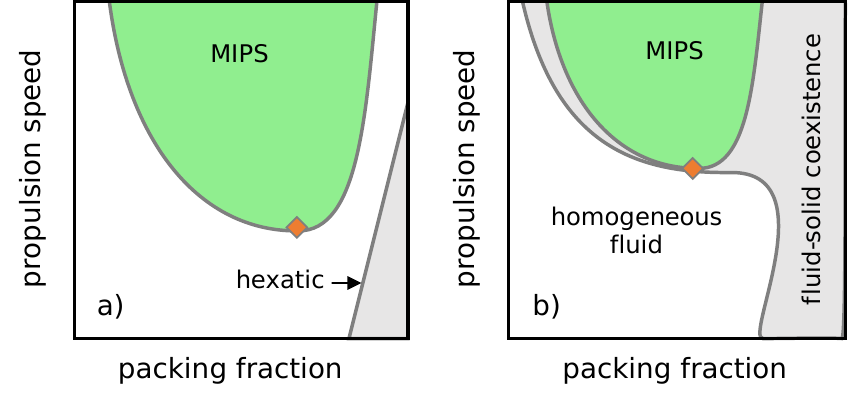}
  \caption{Sketch of the phase diagram of monodisperse active Brownian particles in (a)~two dimensions and (b)~three dimensions. Indicated in green is the two-phase region within which a dense liquid domain coexists with a dilute active gas. The coexistence line ends in a critical point (diamond). Fluid-solid coexistence is indicated in light grey (with the homogeneous solid at even higher densities). In two dimensions there is also a thin hexatic region between fluid and solid~\cite{klamser18}.}
  \label{fig:mips}
\end{figure}

The phase diagram of monodisperse ABPs has been mapped out in simulations for several variants of the model in both two and three dimensions~\cite{redner13,wysocki14,stenhammar14,bialke15,levis17,siebert17,digregorio18,klamser18,turci21,omar21}. In two dimensions, the phase diagram features a coexistence region above a critical speed $\text{Pe}_\text{c}\simeq 40\gg 1$ (in terms of the dimensionless P\'eclet number $\text{Pe}\equiv3v_0\tx/\sig$ with bare propulsion speed $v_0$, orientational correlation time $\tx$, and disk diameter $\sig$), cf. Fig.~\ref{fig:mips}(a). At high densities approaching the close-packing limit, the transition of the passive fluid into a two-dimensional solid continues to non-vanishing P\'eclet numbers with an intermediate hexatic phase~\cite{klamser18}. In three dimensions qualitatively the same behavior is observed~\cite{wysocki14,stenhammar14,siebert17}, although the motility-induced transition is shifted to higher P\'eclet numbers, indicating that in higher dimensions the blocking mechanism is less efficient as one might expect. Very recently it has been shown in simulations that the liquid-gas coexistence in three dimensions is in fact metastable and lies within the two-phase region of gas-solid coexistence~\cite{turci21,omar21}, cf. Fig.~\ref{fig:mips}(b). This is reminiscent of the phase diagram for passive colloidal suspensions with very short-range attractions~\cite{wolde97,anderson02}. The binodal curve of coexisting densities terminates in a critical point that has been investigated numerically in several studies, some of which claim Ising universality~\cite{partridge19,maggi21} as opposed to non-Ising behavior~\cite{siebert18,dittrich21}.

The coexistence of domains of low and high density can be rationalized in field theories for a scalar order parameter $\phi$ related to the density. Cates and coworkers have extensively studied the extension of ``model B''~\cite{hohenberg77} through additional terms of schematic order $\nabla^4$ and $\phi^2$ called ``active model B+''~\cite{stenhammar13,wittkowski14,nardini17,tjhung18,caballero18,cates19}. These additional terms cannot be represented through a free energy functional in general and potentially capture the effects of breaking detailed balance at the particle level. Since these terms are higher derivatives of the field one would expect that they are irrelevant in the renormalization group (RG) sense. However, an explicit one-loop calculation shows that they can induce new fixed points~\cite{caballero18} and thus regions in model space for which the RG flow is pushed away to a non-perturbative regime. To assess this scenario for the various ABP models one would need to explicitly calculate the coefficients of active model B+ from the microscopic model. However, such a systematic link and its verification in numerical simulations is still missing. Another line of investigation has been the mapping to an effective free energy~\cite{paoluzzi16,paoluzzi20} through the unified colored-noise approximation~\cite{jung87}. Such an effective action necessarily yields a $\phi^4$-theory and will not be discussed in the following.


\section{Background}

\subsection{Critical behavior}

Before we review the numerical evidence gathered so far, let us briefly recall some of the basic properties holding close to a critical point. We consider systems in $d$ dimensions that can be described by a scalar order parameter field $\phi(\x)$ that allows to distinguish low density (gas) from high density (liquid), with the average difference vanishing as a power law,
\begin{equation}
  \mean{\phi} \sim (-\tau)^\beta.
  \label{eq:beta}
\end{equation}
Here, $\tau$ measures the distance to the critical point and $\beta$ is the corresponding critical exponent. Other quantities show a similar divergent behavior, in particular the susceptibility
\begin{equation}
  \Int{^d\x'} \mean{\phi(\x)\phi(\x')} \sim |\tau|^{-\gam}
  \label{eq:chi}
\end{equation}
and the correlation length $\xi\sim|\tau|^{-\nu}$ defining two more critical exponents $\gam$ and $\nu$.

In renormalization group (RG) theory, every quantity $x$ comes with a scaling dimension $\Delta_x$ that captures the shift $x\mapsto b^{\Delta_x}x$ under rescaling lengths $\x\mapsto\x/b$ with some factor $b$. Considering the correlation length $\xi\mapsto\xi/b=b^{\Delta_\xi}\xi$, we find $\Delta_\xi=-1$. Scale invariance dictates that the two-point correlation function
\begin{equation}
  \mean{\phi(\x)\phi(\x')} \sim \frac{1}{|\x-\x'|^{d-2+\eta}}
\end{equation}
decays as a power law with ``anomalous'' dimension $\eta$, which yields the scaling dimension
\begin{equation}
  \Delta_\phi = \frac{d-2+\eta}{2}
  \label{eq:sd}
\end{equation}
of the field. The critical exponents imply the relations $\Delta_\xi=-\nu\Delta_\tau=-1$, $\Delta_\phi=\beta\Delta_\tau=\beta/\nu$ from Eq.~\eqref{eq:beta}, and $d-2\Delta_\phi=\gam\Delta_\tau$ from Eq.~\eqref{eq:chi}. The latter can be rearranged to $\gam/\nu=2-\eta$. We thus find the well-known hyperscaling relation
\begin{equation}
  2\Delta_\phi + \gam\Delta_\tau = \frac{2\beta}{\nu} + \frac{\gam}{\nu} = d
  \label{eq:hyper}
\end{equation}
restricting the exponents. Clearly, this relation is fulfilled by the Ising exponents~\cite{onsager44}
\begin{equation}
  \beta = \tfrac{1}{8}, \qquad \gam = \tfrac{7}{4}, \qquad \nu = 1
  \label{eq:ising}
\end{equation}
for $d=2$. However, note that we did not have to invoke a free energy or assume equilibrium-like behavior at any point in order to derive these relations.

\subsection{Numerical results}

\begin{figure}[b!]
  \centering
  \includegraphics{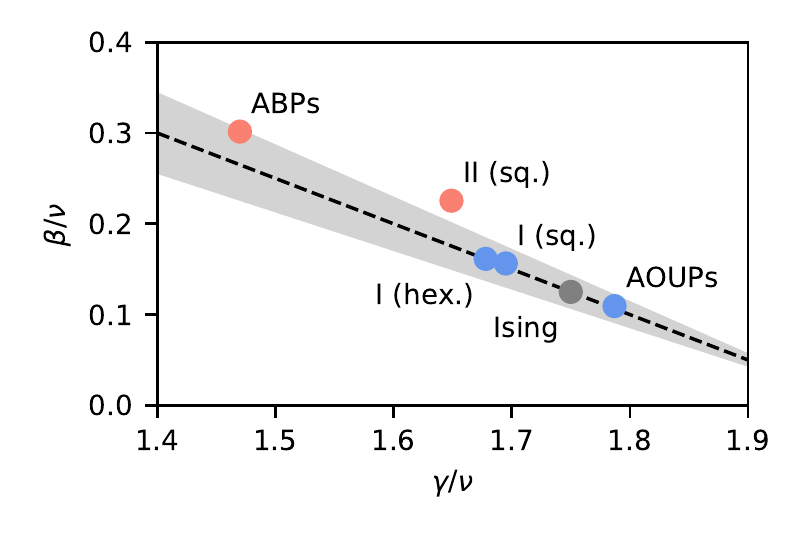}
  \caption{Numerical estimates of critical exponents in two dimensions: off-lattice ABPs~\cite{siebert18} and AOUPs~\cite{maggi21}, and lattice ABPs variants I (on square and hexagonal lattice) and II~\cite{dittrich21}. The dashed line is the hyperscaling relation Eq.~\eqref{eq:hyper} and the gray area indicates an uncertainty of $\pm15\%$. The blue models vary the rotational diffusion time $\tx$ to change the P\'eclet number $\text{Pe}\equiv3v_0\tx/\sig$ whereas the red models change the propulsion speed $v_0$.}
  \label{fig:exp}
\end{figure}

There are now several studies, both on and off lattice, that have determined numerical values for the three critical exponents from finite-size scaling simulations, which are summarized in Fig.~\ref{fig:exp}. Two lattice variants of ABPs have been studied in Ref.~\citenum{dittrich21}. Variant I controls the orientational diffusion at constant speed (basically the model of Partridge and Lee~\cite{partridge19}, who do not report independent values for $\beta$). The difference between square and hexagonal lattice is negligible. We do, however, observe a systematic shift away from the exact Ising values such that the hyperscaling relation Eq.~\eqref{eq:hyper} is still obeyed. Changing the dynamics (on the square lattice) to fix the rotational diffusion and change the propulsion speed leads to variant II~\cite{whitelam18}, which is displaced further from Ising universality. There is an apparent violation of the hyperscaling relation although one has to keep in mind that measuring $\beta$ is challenging in these simulations and its value has the largest uncertainty. Off lattice, ABPs with fixed rotational diffusion and changing the speed $v_0$ have been studied in Ref.~\citenum{siebert18}. The resulting values are even further from Ising universality but still seem to obey the hyperscaling relation within uncertainties. One explanation is that the system sizes studied are insufficient due to large corrections-to-scaling, which is pending further numerical investigations. Also shown are results for active Ornstein-Uhlenbeck particles (AOUPs)~\cite{martin21} obtained by Maggi \emph{et al.}~\cite{maggi21}, which are related to variant I in the sense that the control parameter is the rotational diffusion but now the speed magnitude fluctuates.


\section{Field theory}

\subsection{Adiabatic mean-field theory}

We are interested in an effective field theory in $d$ dimensions that describes the large-scale evolution of the slow collective degrees of freedom, typically the density. Such theories are often constructed from conservation laws and symmetry considerations~\cite{spohn80}. Our starting point here are the hydrodynamic equations for ABPs that follow from the evolution of the one-body joint probability of position and orientation for a tagged particle~\cite{bialke13}. Integrating out the orientation leads to a hierarchy of moments that can be closed (in the large-scale limit, \emph{viz.} vanishing wave vector $q\to0$) after the first non-conserved moment.

The zeroth moment of this hierarchy, the density $\rho(\x,t)$, is conserved, $\partial_t\rho+\nabla\cdot\vec j=0$, with particle current
\begin{eqnarray}
  \vec j = \vec p - \ell^{-2}\nabla\rho + \vec F.
  \label{eq:j}
\end{eqnarray}
The first two terms are the current of non-interacting particles, and the conditional force $\vec F(\x,t)$ takes into account the effect of surrounding particles onto a tagged particle~\cite{speck20}. Throughout we employ dimensionless quantities through measuring time in units of the orientational relaxation time $\tx$ and lengths in units of the persistence length $v_0\tx$ with bare propulsion speed $v_0$. The control parameter thus is the scaled dimensionless persistence length $\ell\equiv v_0\sqrt{\tx/D_0}$ with bare translational diffusion coefficient $D_0$. Moreover, we have scaled the density so that the effective propulsion speed reads $v(\rho)=1-\rho$ and stays zero for $\rho>1$. Such a linear decrease of the speed with density has been confirmed in computer simulations for ABPs~\cite{stenhammar13,fischer19}.

The particle current Eq.~\eqref{eq:j} couples to the local polarization field $\vec p(\x,t)$, the first moment of the orientation, which obeys
\begin{equation}
  \label{eq:p}
  \partial_t\vec p = -\frac{1}{d}\nabla(v\rho) + \ell^{-2}\nabla^2\vec p - \vec p.
\end{equation}
The polarization field is not conserved because of the last term, which captures the relaxation due to the free rotational diffusion. Here we have already neglected the coupling to moments of higher order, which decay even faster and are thus expected to be irrelevant in the $q\to 0$ limit. Neglecting the temporal and spatial derivatives in Eq.~\eqref{eq:p} leads to the adiabatic solution $\vec p_\text{ad}=-\tfrac{1}{d}\nabla(v\rho)$ slaving the polarization to the density. This adiabatic solution together with the current $\vec j_\text{ad}\approx v\vec p_\text{ad}-\ell^{-2}\nabla\rho$ (approximating the force $\vec F\approx-\rho\vec p$~\cite{bialke13}) leads to a Landau mean-field theory with effective free energy~\cite{speck15}
\begin{equation}
  \Psi_\text{ad}[\phi] = \Int{^d\x} \left[\frac{1}{2}\left(\frac{1}{\ell^2}-\frac{1}{8d}\right)\phi^2+\frac{1}{6d}\phi^4\right]
\end{equation}
for the field $\phi=\rho-\rho_\text{c,ad}$. A weakly non-linear analysis around the critical point leads to the same free energy density plus a squared gradient term~\cite{speck14,speck15}. This expansion can be systematically pushed to higher orders~\cite{rapp19,bickmann20}.

For the adiabatic theory, the critical density $\rho_\text{c,ad}=\tfrac{3}{4}$ and critical persistence length $\ell_\text{c,ad}=\sqrt{8d}$ are independent of details of the microscopic interactions. A second route to the mean-field critical point is through the equation of state for the bulk pressure~\cite{solon15,speck21a}
\begin{equation}
  p(\rho;\ell) = \frac{1}{d}\rho(1-\rho) + p_\text{IK} + \ell^{-2}\rho,
\end{equation}
whereby the last term is the ideal gas pressure. For a uniform fluid, the repulsive interactions between particles give rise to an isotropic pressure $p_\text{IK}(\rho;\ell)$ increasing monotonically with density~\cite{solon15,solon18a,speck21a} and conditional force $\vec F=-\nabla p_\text{IK}$. An interaction term of the form $p_\text{IK}=\ell^{-1}f(\rho)$ has been measured for active hard disks ($d=2$) in computer simulations in the homogeneous phase ($\ell<\ellc$)~\cite{solon15,solon18a,speck21a}. The function $f(\rho)$ captures the model-specific interactions. The two conditions for the critical point are $p'(\rhoc;\ellc)=0$ and $p''(\rhoc;\ellc)=0$ (here the prime denotes the derivative with respect to the density). For the data of Ref.~\citenum{solon15}, parametrizing the function $f(\rho)$ one finds $\rhoc\simeq0.69$ reasonably close to $0.75$ but $\ellc\simeq 26.3$ is much larger than the adiabatic prediction $\ell_\text{c,ad}=4$ in $d=2$~\cite{speck21a}.

\subsection{Current expansion}

To improve the adiabatic result and to take into account interparticle forces beyond an effective speed $v(\rho)$, we expand the conditional force in Eq.~\eqref{eq:j}
\begin{equation}
  \vec F \approx -\ell^{-1}\nabla[f_1\delta\rho+f_2(\delta\rho)^2] - \zeta\vec p\delta\rho
  \label{eq:j:exp}
\end{equation}
up to second order in small density fluctuations $\delta\rho\equiv\rho-\rhoc$ away from the critical density $\rhoc$. The two terms inside the brackets stem from the expansion of $f(\rho)$. Even without knowledge of this function, close to the critical point we can find the coefficients from the conditions $p'(\rhoc)=0$ and $p''(\rhoc)=0$ at $\ell=\ellc$, which imply $f_2=\ellc/d$ and $f_1=\ellc(\al-\ellc^{-2})$. Here we have introduced the parameter $\al>\ellc^{-2}>0$ determining the critical density through $\rhoc=\tfrac{1}{2}+\tfrac{d}{2}\al$. The values for $\ellc$ and $\al$ are determined by the functional form of $f(\rho)$ but are considered an input to the theory in the following.

The last term in Eq.~\eqref{eq:j:exp} takes into account the coupling between polarization and density fluctuations, which is absent in the uniform fluid. We thus obtain the coupled hydrodynamic equations
\begin{equation}
  \label{eq:p0}
  \partial_t\delta\rho = c\nabla^2\delta\rho - \nabla\cdot\vec p + \zeta\nabla\cdot(\vec p\delta\rho) + \zeta_2\nabla^2(\delta\rho)^2
\end{equation}
and
\begin{equation}
  \label{eq:pp}
  \partial_t\vec p = \ell^{-2}\nabla^2\vec p - \vec p + \al\nabla\delta\rho + \frac{1}{d}\nabla(\delta\rho)^2
\end{equation}
with coefficients $c\equiv f_1/\ell+\ell^{-2}$ and $\zeta_2\equiv\ellc/(d\ell)$.

\subsection{Linear analysis}
\label{sec:lin}

We collect the density and polarization into $d+1$ fields $p_i(\x,t)$, whereby $p_0=\delta\rho$. In Fourier space
\begin{equation}
  p_i(\x,t) = \IntF{\vec q} e^{\im\vec q\cdot\x} p_i(\vec q,t)
\end{equation}
with magnitude $q=|\vec q|$ of the wave vector $\vec q$. The linearized hydrodynamic equations can be written $\partial_tp_i=L_{ij}p_j$ with $(d+1)\times(d+1)$ matrix
\begin{equation}
  \label{eq:L}
  \vec L(\vec q) =
  \left(\begin{array}{cccc}
    -c q^2 & -\im q_1 & -\im q_2 & \cdots \\
    \al\im q_1 & -(q/\ell)^2-1 & 0 \\
    \al\im q_2 & 0 & -(q/\ell)^2-1 \\
    \vdots & & & \ddots
  \end{array}\right).
\end{equation}
We follow the summation convention and sum over repeated indices. It is easy to see that the eigenvalues of this matrix are $-1+\mathcal O(q^2)$ and thus negative except for the largest eigenvalue $\sig_+(q)$, which can be expanded for small $q$ into $\sig_+(q)\approx-aq^2-\kap q^4$ with the two coefficients
\begin{equation}
  a = c - \al \approx (\al+\ellc^{-2})(1-\ell/\ellc)
\end{equation}
and ``stiffness'' $\kap=\al(1-a)\approx\al$. For $a<0$ ($\ell>\ellc$) the eigenvalue $\sig_+$ becomes positive for small $q$, signaling a dynamic instability.

The corresponding eigenvector $L_{ij}v^+_j=\sig_+v^+_i$ reads
\begin{equation}
  \vec v^+(q) \approx \left(\begin{array}{c}
    1+(1-a)q^2 \\ \al\im q_1 \\ \al\im q_2 \\ \vdots
  \end{array}\right)
\end{equation}
up to order $q^2$. So far we have ignored that the hydrodynamic equations do come with a noise term, both due to the influence of the solvent and due to the coarse-graining into fields. The noise is modeled as Gaussian with zero mean and correlations
\begin{equation}
  \mean{\eta(\hat q)\eta(\hat q')} = 2D q^2(2\pi)^{d+1}\delta^d(\vec q+\vec q')\delta(\om+\om').
  \label{eq:K}
\end{equation}
Note the factor $q^2$, which appears due to the conservation of the density. The evolution of the eigenmode $\vec v^+$ with coefficient $\phi(\vec q,\om)$ is thus $\phi=G_0\eta$ with free propagator
\begin{equation}
  G_0(\hat q) = G_0(q,\om) \equiv \frac{1}{-\im\om + a q^2 + \kap q^4}
  \label{eq:G0}
\end{equation}
after switching to the frequency domain and writing $\hat q\equiv(\vec q,\om)$. The correlation function of the eigenmode becomes
\begin{equation}
  \mean{\phi(\hat q)\phi(\hat q')} = (2\pi)^{d+1}C_0(\hat q)\delta^{d+1}(\hat q+\hat q')
\end{equation}
with $C_0(\hat q)\equiv 2Dq^2G_0(\hat q)G_0(-\hat q)$. The static structure factor follows as
\begin{equation}
  S_0(q) = \frac{1}{2\pi}\IInt{\om}{-\infty}{\infty} C_0(q,\om) = \frac{D}{a+\kap q^2},
\end{equation}
from which we read off the correlation length $\xi=\sqrt{\kap/a}\propto a^{-\nu}$ with exponent $\nu=\tfrac{1}{2}$. The correlation length diverges at the critical point with $S_0(q)\sim q^{-2}$. In general, $S(q)\sim q^{-2+\eta}$, and the two critical exponents $\nu=\tfrac{1}{2}$ and $\eta=0$ define the Gaussian fixed point.

\subsection{Non-linear theory: Eigenmode expansion}

We return to the evolution equation Eq.~\eqref{eq:p0} for the density field $p_0$, which now reads (with index $k=1,\dots,d$)
\begin{multline}
  -\im\om p_0 = L_{0i}p_i + \im\zeta q_k\int_{\hat k}p_k(\hat k)p_0(\hat q-\hat k) \\ - \zeta_2q^2\int_{\hat k}p_0(\hat k)p_0(\hat q-\hat k)
  \label{eq:p0:fourier}
\end{multline}
writing $\hat k\equiv(\vec k,\Om)$ and
\begin{equation}
  \int_{\hat k} \equiv \int\frac{\dd^d\vec k\dd\Om}{(2\pi)^{d+1}}.
\end{equation}
Clearly, the density couples to the polarization fields $p_k$ through the first non-linear term. For a closure, we expand the fields into the eigenvectors of the matrix $\vec L$,
\begin{equation}
  \label{eq:eig}
  p_i = \sum_{n=0}^d \phi^{(n)} v_i^{(n)} \approx \phi v^+_i.
\end{equation}
In the following, we restrict our attention to the dynamics of the slowest mode amplitude $\phi(\vec q,t)$ corresponding to the eigenvalue $\sig_+(q)$. Inserting $p_0=v^+_0\phi$ and $p_k=\al\im q_k\phi$ into Eq.~\eqref{eq:p0:fourier} yields
\begin{multline}
  -\im\om v^+_0\phi = \sig_+v^+_0\phi - \int_{\hat k} [\zeta\al\vec q\cdot\vec k v^+_0(\vec q-\vec k) \\ +\zeta_2q^2v^+_0(k)v^+_0(\vec q-\vec k)] \phi(\hat k)\phi(\hat q-\hat k).
\end{multline}
Gathering terms into the free propagator $G_0$ [Eq.~\eqref{eq:G0}] and including again the noise field leads to the evolution equation
\begin{equation}
  \phi = G_0\eta + G_0\int_{\hat k}v_2(\vec k,\vec q-\vec k)\phi(\hat k)\phi(\hat q-\hat k),
  \label{eq:phi}
\end{equation}
which is our first central result. It suggests that the scalar order parameter for the critical fluctuations is not exactly the density $\delta\rho$ but the coefficient $\phi$ of the slowest mode with density $\delta\rho\approx\phi-(1-a)\nabla^2\phi$ and polarization field $\vec p\approx\al\nabla\phi$ in real space.

In Eq.~\eqref{eq:phi} we have defined the bare symmetric vertex 
\begin{equation}
  v_2(\vec k_1,\vec k_2) = \frac{1}{2}[v'_2(\vec k_1,\vec k_2)+v'_2(\vec k_2,\vec k_1)]
\end{equation}
with
\begin{equation}
  \label{eq:v2}
  v'_2(\vec k_1,\vec k_2) \equiv -v^+_0(k_2)[\zeta\al\vec q\cdot\vec k_1+\zeta_2q^2v^+_0(k_1)]/v^+_0(q).
\end{equation}
Since there is only one field $\phi$, $v_2(\vec k_1,\vec k_2)$ must be symmetric with respect to exchanging $\vec k_1\leftrightarrow\vec k_2$. Note the effective ``momentum conservation'' $\vec q=\vec k_1+\vec k_2$. In real space, this function represents a non-local kernel and upon expansion generates spatial derivatives of $\phi$.

Let us see what these derivatives are. At the mean-field critical point with $a=0$ we insert $v^+_0(q)\approx 1+q^2$. To lowest order $q^2$ we find
\begin{equation}
  v_2^{(2)} = -\frac{1}{2}\zeta\al\vec q\cdot(\vec k_1+\vec k_2) - \zeta_2q^2 = -\bar bq^2
\end{equation}
with $\bar b=\zeta\al/2+\zeta_2$, which corresponds to $\bar b\nabla^2\phi^2$ in real space. The next order comprises products of four wave vectors,
\begin{equation}
  v_2^{(4)} = \bar bq^4 - \zeta_2q^2(k_1^2+k_2^2) - \frac{1}{2}\zeta\al(\vec q\cdot\vec k_1 k_2^2+\vec q\cdot\vec k_2 k_1^2) .
\end{equation}
The second term is split using $k_1^2+k_2^2=q^2-2\vec k_1\cdot\vec k_2$ so that
\begin{equation}
  v_2^{(4)} = \frac{c_1}{2}q^4 + c_2 q^2\vec k_1\cdot\vec k_2 - \frac{c_3}{2}(\vec q\cdot\vec k_1 k_2^2+\vec q\cdot\vec k_2 k_1^2)
\end{equation}
with three further coefficients (using $\zeta_2=1/d$)
\begin{equation}
  c_1 = 2(\bar b - \zeta_2) = \zeta\al, \quad c_2 = 2\zeta_2 = \frac{2}{d}, \quad c_3 = \zeta\al.
  \label{eq:c}
\end{equation}
In real space, these terms correspond to $\frac{1}{2}c_1\nabla^4\phi^2$, $c_2\nabla^2|\nabla\phi|^2$, and $-c_3\nabla\cdot[(\nabla^2\phi)\nabla\phi]$. The values for the bare coefficients $\bar b$ and $c_n$ are given to leading order at the mean-field critical point but in principle they depend on the distance $a$ from the critical point.

The evolution equation~\eqref{eq:phi} for the order parameter field $\phi(\x,t)$ in real space thus becomes
\begin{multline}
  \partial_t\phi = \nabla^2(a\phi + \bar b\phi^2 - \kap\nabla^2\phi) + \frac{c_1}{2}\nabla^4\phi^2 + c_2\nabla^2|\nabla\phi|^2 \\ - c_3\nabla\cdot[(\nabla^2\phi)\nabla\phi] + \eta,
  \label{eq:amb}
\end{multline}
which is precisely active model B+ and constitutes our second central result. It includes all quadratic terms involving up to four $\nabla$'s that respect translational and rotational symmetry. Clearly, the coupling $\zeta$ between polarization and density in the conditional force [Eq.~\eqref{eq:j:exp}] is crucial to generate the non-linear terms with coefficients $c_1$ and $c_3$. Equation~\eqref{eq:amb} corresponds to a conserved dynamical equation for $\phi$ with current
\begin{equation}
  \vec j_\phi = -\nabla\fd{\Psi}{\phi} - \nabla\left(\frac{c_1}{2}\nabla^2\phi^2 + c_2|\nabla\phi|^2\right) + c_3(\nabla^2\phi)\nabla\phi
  \label{eq:j:phi}
\end{equation}
involving the Landau-Ginzburg functional
\begin{equation}
  \Psi[\phi] \equiv \Int{^d\x} \left[\frac{\kappa}{2}|\nabla\phi|^2+\frac{a}{2}\phi^2+\frac{\bar b}{3}\phi^3+\frac{u}{4}\phi^4\right].
  \label{eq:lg}
\end{equation}
Here we have added a $\phi^4$ term with coefficient $u>0$, which would have been generated by our expansion of $f(\rho)$ to third order [Eq.~\eqref{eq:j:exp}] (and neglecting all other terms that would appear at this order). The cubic term $\bar b\phi^3$ can be eliminated through shifting $\phi\mapsto\phi+\phi_0$ with $\phi_0=-\bar b/(3u)$. This shift leads to a linear term $\propto\phi$ in $\Psi$ which does not change the evolution equation. It also reduces $a\mapsto a-\bar b^2/(3u)$ with the Gaussian critical point still defined by the condition $a=0$.


\section{Discussion}

\subsection{Coexistence}
\label{sec:coex}

Before going to the critical point let us look at state points $\ell>\ellc$ in the two-phase region with coexisting dense and dilute domains. First, note that the terms involving $\kap$, $c_1$, and $c_2$ in Eq.~\eqref{eq:amb} can be rewritten as
\begin{equation}
  \nabla^2\left[-(\kap-c_1\phi)\nabla^2\phi + \lam|\nabla\phi|^2\right]
\end{equation}
with $\lam\equiv c_1+c_2$. If we compare this expression with the functional derivative
\begin{equation}
  \fd{}{\phi}\Int{^d\x} \frac{1}{2}\kap(\phi)|\nabla\phi|^2 = -\kap(\phi)\nabla^2\phi - \frac{1}{2}\kap'(\phi)|\nabla\phi|^2
\end{equation}
we see that with $\kap(\phi)\equiv\kap-c_1\phi$ and $\lam=-\kap'(\phi)/2$ these terms can be absorbed into the Landau-Ginzburg functional Eq.~\eqref{eq:lg}. For the non-linear coefficients $c_n$ this leads to the conditions $c_3=0$ and $c_1+2c_2=0$ along which a free energy functional is restored implying equilibrium-like behavior.

\begin{figure}[t]
  \centering
  \includegraphics{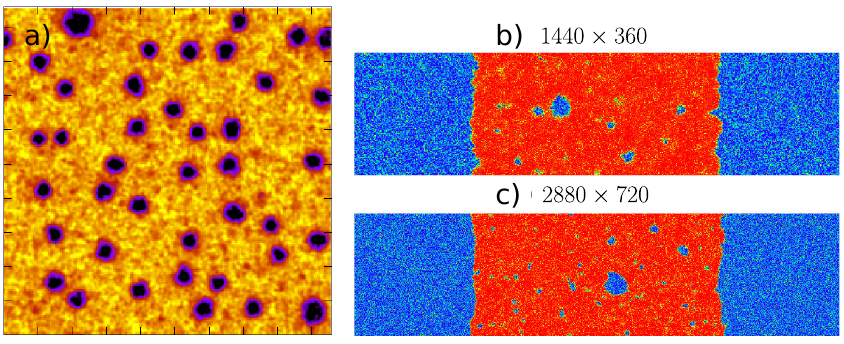}
  \caption{Numerical results at coexistence. (a)~Numerical solution of active model B+ [Eq.~\eqref{eq:amb}] for $c_1=0$, $c_2=1$, and $c_3=4$ inside the two-phase region. [Reproduced from Ref.~\citenum{tjhung18}] (b,c)~Simulation snapshots for active Brownian particles with anisotropic mobility tensor for two system sizes. [Reproduced with permission from Ref.~\citenum{shi20}]}
  \label{fig:num}
\end{figure}

Moving away from the equilibrium-like regime with $c_3\neq 0$, Eq.~\eqref{eq:amb} predicts a qualitatively new feature: droplets of the (local) minority phase have a typical finite size with smaller droplets growing and larger droplets shrinking~\cite{tjhung18}. This reverse Ostwald ripening leads to both ``bubbly phase separation'' with gas droplets suspended within the dense liquid as well as a state that resembles microphase separation [Fig.~\ref{fig:num}(a), see also Ref.~\citenum{thomsen21} for the emergence of periodic solutions in a related type of field theory]. This process implies a distribution of droplet sizes that is peaked around the typical size as confirmed in numerical solutions of Eq.~\eqref{eq:amb}. At variance with this picture, the distribution of droplet sizes in simulations of ABPs decays algebraically and scales with the system size~\cite{shi20}, cf. Fig.~\ref{fig:num}(b,c) (the authors even had to introduce an anisotropic mobility tensor to promote gas bubbles). Indeed, a closer looks suggests that the creation of these bubbles is connected to defects between domains of differently aligned hexatic order~\cite{caporusso20}. Moreover, it has been shown recently that the coexisting densities predicted from analytical expressions for $\kap(\phi)$ and $\lam(\phi)$, which have been derived from the stress, are in excellent agreement with numerical results for ABPs in $d=2$ setting $c_3=0$~\cite{speck21a}. Together, these results suggest that the coexistence of (standard) ABPs is closer to active model B with $c_3=0$.

\subsection{Perturbative renormalization}

We now turn to the large-scale behavior in the vicinity of the critical point. The scaling dimension $\Delta_x$ determines whether the quantity $x$ is relevant ($\Delta_x>0$) or irrelevant ($\Delta_x<0$) with the borderline $\Delta_x=0$ termed marginal. In dimensions $d>4$ the non-linear terms in Eq.~\eqref{eq:amb} become irrelevant (they are mapped to zero under the RG flow) and the critical behavior is determined by the Gaussian critical fixed point (see Sec.~\ref{sec:lin}) with $a$ the only relevant parameter determining the distance to the Gaussian critical point. The accepted picture for the Ising model is that at $d_\text{c}=4$ another fixed point, the Wilson-Fisher fixed point (WF), detaches from the Gaussian fixed point (G). WF moves away as $d$ is reduced and for $d<4$ determines the critical behavior (G now is repulsive along $u$ while WF is attractive so that any initial point not fine-tuned to $a=0$ will flow into WF). The WF has been studied extensively through perturbative RG with good quantitative agreement with the critical exponents of the Ising model in $d=2$ and (numerically determined) in $d=3$~\cite{leguillou85}.

Wilson's shell renormalization scheme produces flow equations for the model parameters $\vec x=(\kap,D,a,\dots)$ of the form
\begin{equation}
  \partial_lx_n = (\Delta_n + \psi_n)x_n.
  \label{eq:rg:flow}
\end{equation}
Here, $\psi_n(\vec x)$ denotes the graphical corrections due to the recursive solution of Eq.~\eqref{eq:phi}, which generates a series of terms with increasing powers of the vertices. In a nutshell, internal wave vectors are integrated out in the thin shell $\Lam/(1+\delta l)<k<\Lam$ with microscopic cut-off $\Lam$ leading to $\tilde x_n=(1+\psi_n\delta l)x_n$. Restoring the cut-off requires to scale external wave vectors $\vec q\to b\vec q$ with $b=1+\delta l$ implying $x'_n=b^{\Delta_n}\tilde x_n\approx(1+\Delta_n\delta l)(1+\psi_n\delta l)x_n$, leading to Eq.~\eqref{eq:rg:flow} in the limit $\delta l\to0$. The model parameters $\vec x(l)$ become a function of the scale $b=e^l$. Since $\xi(l)=\xi_0e^{-l}$ any initial values $\vec x_0$ with \emph{finite} correlation length $\xi_0$ will flow to $\xi\to0$ as $l\to\infty$. Only when hitting $\xi\to\infty$ will the flow take us to a critical fixed point of the RG flow.

Caballero \emph{et al.} have calculated Wilson's flow equations for active model B+ including the one-loop corrections~\cite{caballero18}. Good references for technical details of dynamic RG calculations are Refs.~\citenum{medina89,tauber12}. For completeness, let us recapitulate the main points. First, note that the naive scaling dimensions keeping Eq.~\eqref{eq:amb} invariant under rescaling length and time are $\Delta_\kap=z-4$, $\Delta_u=z-2-2\Delta_\phi$, and $\Delta_D=\Delta_\kap+\eta=z-2-d+2\Delta_\phi$ using Eq.~\eqref{eq:sd} with dynamic exponent $z$. Moreover, the coefficients $c_n$ have the same scaling dimension $\Delta_c=z-4-\Delta_\phi$. Second, we introduce suitable reduced parameters
\begin{equation}
  \bar u \equiv \frac{uD}{\kap^2}K_d\Lam^{4-d}, \quad 
  \bar c_n \equiv \frac{c_nD^{1/2}}{\kap^{3/2}}K_d^{1/2}\Lam^{d/2-1}
\end{equation}
that remove the dependence on the unknown exponents $z$ and $\Delta_\phi$ with $K_d\equiv S_d/(2\pi)^d$ ($S_d$ is the surface area of a hypersphere in $d$ dimensions).

\begin{figure}[t]
  \includegraphics{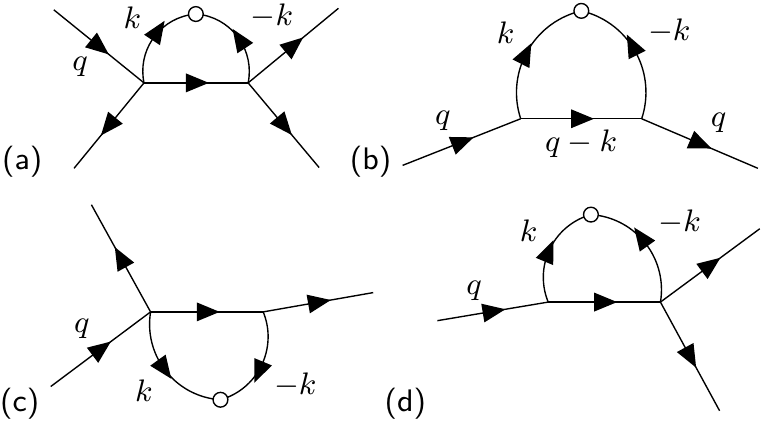}
  \caption{Graphical representation of the relevant one-loop corrections with two vertices. Outgoing external legs represent $\phi$ [noise in (b)], internal lines the free propagator $G_0$ [Eq.~\eqref{eq:G0}], and the open circle is the correlation function $C_0$. Each edge has an orientation and carries a wave vector (and frequency) such that at each vertex $v_n(\vec k_1,\dots,\vec k_n)$ the sum of outgoing wave vectors $\vec k_i$ equals the single ingoing wave vector (orientations indicated by arrows).}
  \label{fig:graphs}
\end{figure}

To complement the discussion of Ref.~\citenum{caballero18} and relate to the numerical results for the critical exponents of ABPs, let us focus on the RG flow close to two dimensions through setting $d=2+\eps$ with $\eps$ assumed to be small. To lowest order of the non-linear coefficients $u$ and $c_n$, only graphs with two vertices contribute, which are shown in Fig.~\ref{fig:graphs}: Fig.~\ref{fig:graphs}(a) is the conventional Ising contribution $\psi_u\propto u^2$, Fig.~\ref{fig:graphs}(b) leads to the correction $\psi_\kap\propto c_n^2$, and Figs.~\ref{fig:graphs}(c,d) give rise to corrections $\psi_{c_n}\propto uc_n$. The final flow equations read
\begin{equation}
  \partial_l\bar u = (2-\eps-9\bar u-2\psi_\kap)\bar u
  \label{eq:rg:u}
\end{equation}
and
\begin{align}
  \label{eq:rg:c1}
  \partial_l\bar c_1 &= \left(-\frac{\eps}{2}-\frac{3}{2}\psi_\kap\right)\bar c_1 - 3\bar u(\bar c_1+2\bar c_2) + \frac{9}{2}\bar u\bar c_3, \\
  \label{eq:rg:c2}
  \partial_l\bar c_2 &= \left(-\frac{\eps}{2}-\frac{3}{2}\psi_\kap\right)\bar c_2 + \frac{3}{2}\bar u\bar c_1\eps - 3\bar u\bar c_3, \\
  \label{eq:rg:c3}
  \partial_l\bar c_3 &= \left(-\frac{\eps}{2}-\frac{3}{2}\psi_\kap\right)\bar c_3.
\end{align}
Here we have used that the noise strength $D$ does not receive corrections ($\psi_D=0$) and we have expanded $\psi_{c_n}$ with the coefficients taken from Ref.~\citenum{caballero18} to leading order of $\eps$.

For $c_n=0$ implying $\psi_\kap=0$ we immediately recover the WF fixed point at $\bar u^\ast=(2-\eps)/9$. Of course, the exact value is not to be trusted close to two dimensions but it is widely accepted that the qualitative picture is correct and the WF corresponds to Ising criticality even away from the upper critical dimension $d_\text{c}=4$. For the following discussion it is sufficient to assume that the WF sits at some positive $\bar u^\text{WF}\sim\mathcal O(1)$.

\begin{figure}[t]
  \centering
  \includegraphics{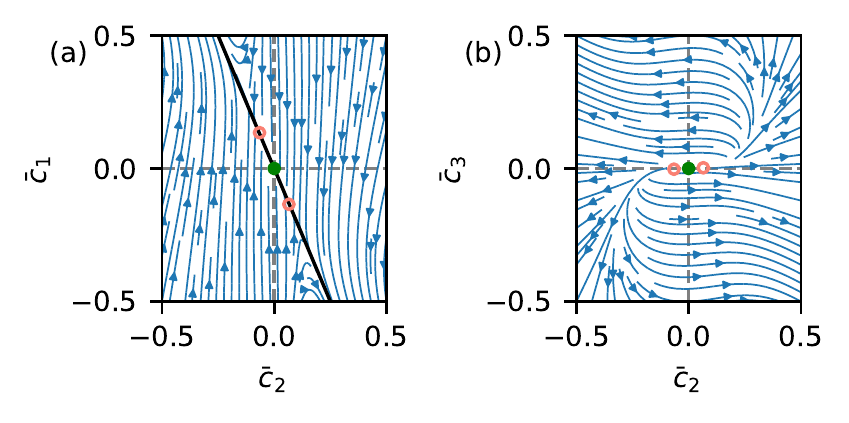}
  \caption{Renormalization flow portrait. (a)~Varying $\bar c_1$ and $\bar c_2$ for $\bar c_3=0$ and $\bar u^\ast=(2-\eps)/9$ at $\eps=0.04$. The filled symbol indicates the Wilson-Fisher fixed point and the open symbols the fixed points Eq.~\eqref{eq:rg:f4}. Along the solid black line $\bar c_1+2\bar c_2=0$ corresponding to an equilibrium-like free energy functional. (b)~Varying $\bar c_2$ and $\bar c_3$ with $\bar c_1+2\bar c_2=0$.}
  \label{fig:flow}
\end{figure}

From Ref.~\citenum{caballero18}, we find
\begin{equation}
  \psi_\kap = -\frac{1}{2}\bar c_1^2 + \bar c_1\bar c_2 - \frac{1}{4}\bar c_1\bar c_3 + \bar c_2^2 - 2\bar c_2\bar c_3
\end{equation}
at $d=2$ (which we have confirmed independently). To determine further fixed points, we thus have to solve the set of cubic equations. The first pair is determined by $\psi_\kap=-\eps/3$ which fixes $\bar u^\ast=(2-\eps/3)/9$. The remaining equations are independent of $\bar u^\ast$ with solutions
\begin{equation}
  \bar c_1^\ast = \pm\frac{2}{3}\eps^{1/2}, \quad \bar c_2^\ast = \mp\frac{1}{3}\eps^{1/2}, \quad \bar c_3^\ast = \pm\frac{1}{3}\eps^{3/2}
  \label{eq:rg:f4}
\end{equation}
to leading order consistent with the perturbative expansion of Eq.~\eqref{eq:amb}. These fixed points merge with the WF for $\eps\to0$. Moreover, they lie along the line $\bar c_1+2\bar c_2$ corresponding to an equilibrium-like free energy functional with stiffness $\kap(\phi)$ discussed in the previous section~\ref{sec:coex}. Further fixed points lie outside the perturbative region. Figure~\ref{fig:flow} visualizes the flow of the non-linear coefficients around the fixed points.

As one might expect, the scaling dimensions of the WF are not changed by the new fixed points~\eqref{eq:rg:f4}. However, these fixed points strongly limit the region of the non-linear parameters for which the WF is attractive, which vanishes entirely as $\eps\to0$. The new fixed points are attractive along the directions $\bar u$ and $\bar c_1$, and repulsive along $\bar c_3$. At least for small $\eps$, the flow is attracted to the line $\bar c_1+2\bar c_2=0$ running away from the perturbative region outside the new fixed points. For $\eps=0$ the linearized flow around these fixed points reads
\begin{equation}
  \partial_l\delta\bar{\vec c} = -3\bar u
   \left(\begin{array}{ccc}
      1 & 2 & -\tfrac{3}{2} \\
      0 & 0 & 1 \\
      0 & 0 & 0
    \end{array}\right)\delta\bar{\vec c},
\end{equation}
which has eigenvector $\vec e_1=(1,0,0)^T$ with eigenvalue $\Delta_1=-3\bar u$ and eigenvector $\vec e_0=(-2,1,0)^T$ with zero eigenvalue. Figure~\ref{fig:flow}(a) shows that the flow along $\vec e_0$ is repulsive and thus marginally relevant.


\section{Conclusions}

As a step towards a comprehensive link between microscopic particle-based models of scalar active matter and their continuum field theories, we have presented a new route to active model B+ [Eq.~\eqref{eq:amb}] starting from the intermediate hydrodynamic equations for density [Eq.~\eqref{eq:p0}] and polarization [Eq.~\eqref{eq:pp}]. We have identified the coupling $\zeta$ between density fluctuations and polarization in the quadratic expansion of the conditional force [Eq.~\eqref{eq:j:exp}] as the crucial term that generates the non-potential contributions to Eq.~\eqref{eq:amb}. Still missing is a strategy to calculate $\zeta$ from particle-based simulations of the microscopic equations of motion.

We have then revisited the RG flow close to two dimensions. The picture that follows from the perturbative RG in $d\to2$ is the following: While the naive scaling dimension $\Delta_c=-\eps/2$ of the non-linear coefficients $c_n$ indicates that they are irrelevant, they do give rise to a pair of new fixed points that destabilize the WF in two dimensions. Consequently, the parameters $c_n$ run off to a non-perturbative regime. While one can only speculate about their fate, it is interesting to note that $\bar c_1$ and $\bar c_2$ are pulled towards values that restore the Landau-Ginzburg functional [Eq.~\eqref{eq:lg}] with field-dependent stiffness $\kap(\phi)=\kap-c_1\phi$ and current
\begin{equation}
  \vec j_\phi = -\nabla\fd{\Psi}{\phi} + c_3(\nabla^2\phi)\nabla\phi.
\end{equation}
This suggests that two outcomes are possible: for $\zeta=0$ implying $c_3=0$ one would expect indeed equilibrium-like behavior (with a strong-coupling equilibrium fixed point residing one the line $c_1+2c_2=0$). If $\zeta\neq 0$ and the coupling between density and polarization in Eq.~\eqref{eq:j:exp} is relevant then $c_3$ runs off from its initial value $c_3(0)=\zeta\al$ and it is conceivable that the critical behavior is governed by a non-Ising fixed point. The strong discrepancy between the numerical results for different models (Fig.~\ref{fig:exp}) hints at such a possibility, in particular between (on and off-lattice) ABPs and AOUPs.

Maggi \emph{et al.} have studied the fluctuation-dissipation theorem in AOUPs in $d=2$ dimensions as a function of wave vector $q$~\cite{maggi21a}. They find that the FDT is fulfilled on large length scales (small $q$) but violated on shorter scales. This prompts them to replace the white-noise Eq.~\eqref{eq:K} by colored noise characterized by a spatial and temporal decay scale. However, in the limit $q\to0$ such colored noise will again reduce to white noise and does not affect the critical properties. It does agree with our observation that the equilibrium line is attractive, and at least for AOUPs on large scales equilibrium-like behavior obeying the fluctuation-dissipation theorem is restored. It would be interesting to study if that extends to ABPs with fixed speed magnitude, which are also characterized by the absence of persistent particle currents in the steady state. Further numerical studies are certainly welcome to shed light on the critical behavior of active particles.


\begin{acknowledgments}
  This work has been supported by the Deutsche Forschungsgemeinschaft within the CRC TRR 146 (grant no. 233630050, project C7). Stimulating discussions with P. Virnau on the numerical results are gratefully acknowledged.
\end{acknowledgments}


%

\end{document}